\documentclass[conference]{IEEEtran}
\usepackage{amsmath,amssymb}
\usepackage{bm}
\usepackage[dvipdfmx]{color}
\usepackage[dvipdfmx]{graphicx}

\graphicspath{{figure/}}

\newtheorem{theorem}{Theorem}
\newtheorem{remark}{Remark}

\makeatletter
\let\MYcaption\@makecaption
\makeatother
\usepackage[font=footnotesize]{subcaption}
\makeatletter
\let\@makecaption\MYcaption
\makeatother

\usepackage{setspace}

\begin{document}
\title{
Analysis of Breakdown Probability of Wireless Sensor Networks with Unreliable Relay Nodes }
\author{
  \IEEEauthorblockN{Takayuki Nozaki,}
  \IEEEauthorblockA{Yamaguchi University, Japan\\
      Email: tnozaki@yamaguchi-u.ac.jp} 
  \and 
  \IEEEauthorblockN{Takafumi Nakano, and Tadashi Wadayama}
  \IEEEauthorblockA{Nagoya Institute of Technology,   Japan\\
      Email: wadayama@nitech.ac.jp} 
}

\maketitle

\begin{abstract}
In the present paper,  
we derive an upper bound of the average network breakdown probability of packet networks with unreliable relay nodes.
We here assume that relay nodes get independently broken with a given node breakdown probability.
A survivor graph is the induced subgraph obtained by removing the broken relay nodes and their connecting edges from the original graph.
If the survivor network is disconnected, we consider a network breakdown happens.
The primal contribution of the paper is to derive an upper bound of the average network breakdown probability, where the expectation is taken over a regular graph ensemble.
The proof of the bound is based on a natural one-to-one correspondence between a regular graph and a regular bipartite graph, 
and also on enumeration of bipartite graphs satisfying certain conditions.
This proof argument is inspired by the analysis of weight distribution for low-density parity-check codes.
Compared with estimates of the average network breakdown probability obtained by computer experiments, 
it is observed that the upper bound  provides the values 
which are not only upper bounds but also precise estimates of 
the network breakdown probability when the node breakdown probability is small.
\end{abstract}

\section{Introduction}\label{sec:intro}

Wireless sensor networks (WSN) regain huge interest from both academia and industry in the context of the development of IoT (Internet of Things). 
It is expected that WSN can provide tremendous applications in the near future such as security surveillance, 
environmental monitoring, disaster monitoring,  factory monitoring, and so on.
In many cases, sensor nodes have short-range wireless communication connectivity and packet relay capability in addition to its sensing capability.  
The sensor data are aggregated by gateways and a packet-based relaying protocol is commonly 
used to convey the data from sensors to the gateway. 
Sensor nodes are deployed in a large area which may not be an ideal environment to place small electronic devices.
In some situations, sensor nodes are directly exposed to severe environments, e.g., outdoor environment, 
and it may continuously cause damage to the sensor nodes.
Another source of unreliability is limited batteries of sensor nodes that are not easy to recharge.
Sufficient maintenance to a number of sensor nodes may be hopeless in some situations because of its huge cost.
This means that we need to take care of possible breakdowns,  malfunctions,  and flat batteries of sensor nodes for designing and operating WSN.

A natural abstraction of WSN is to exploit an undirected graph to represent the network.
A node corresponds to a sensor node and an edge corresponds to a wireless link connecting two sensor nodes. 
Connectivity of such graphs is of prime importance because it ensures successful packet-based communications over the network
with appropriate routing.
If any two distinct nodes in a graph have a path connecting both nodes, the graph is said to be connected. 
The paper by Li et al.~\cite{Li} gives a survey of the works regarding the connectivity of several random graph classes.

The problem to evaluate robustness of networks in terms of the connectivity 
under the assumption of some edge breakdowns  is known as the {\em network reliability problem}.
The origin of the network reliability problem is 
the celebrated paper by Moore and Shannon \cite{REL} which dates back to 50's.
Since then, the topic has been extensively studied primarily in the fields of theoretical computer science.
Evaluation of the network breakdown probability for a given undirected graph is known to be a computationally hard problem. 
Provan and Ball \cite{NP1} and Valiant \cite{NP2} showed that the network reliability problems are $\#{\cal P}$-complete,
which is a complexity class at least as intractable as ${\cal NP}$. 
Karger \cite{Poly} presented a randomized polynomial time approximation algorithm for the all terminal network reliability problems.

In this paper, we discuss the problem to evaluate the probability that a random network becomes disconnected 
under the assumption of  {\em node breakdowns} (i.e., relay node breakdowns) instead of edge breakdowns
 assumed in the literatures of the network reliability problem.
The main motivation of our work is to establish a solid theoretical foundation to analyze WSN with {\em unreliable relay nodes}. 
Although breakdowns of relay nodes degrade the system performance,  they may not be devastating if the network retains its connectivity.
From this respect, the robustness of WSN  can be evaluated by measuring the probability of events that
the network becomes disconnected. One may be able to use such information to optimize certain parameters of WSN under the cost constraint.
The problem setup introduced in this paper appears simple and natural but it has not been seriously studied as far as the authors know.

We will present a probabilistic analysis of connectivity of regular random graphs with relay node breakdowns in this paper.
We here assume that relay nodes get independently broken with the node breakdown probability $\epsilon$.
The remaining graph, called a survivor graph in this paper, is the induced subgraph obtained 
by removing the broken relay nodes and their connecting edges from the original graph.
If the survivor network is disconnected, we consider a {\em network breakdown} happens.

The primal contribution of the paper is to derive an upper bound of the expected network breakdown probability.
The network breakdown probability is the probability such that a randomly chosen network is disconnected under the assumption of probabilistic breakdowns of relay nodes.
The expectation is taken over a $\lambda$-regular graph ensemble.
The upper bound indicates the typical behavior of the network breakdown probability.
The proof argument of the bound is inspired by the analysis of weight distribution for low-density parity-check (LDPC) codes.
In order to derive the bound, we introduced a special class of bipartite graphs that is similar to the Tanner graphs \cite{modern} of LDPC codes.
Counting the number of bipartite graphs satisfying certain conditions leads to the main result.
The argument is similar to the ensemble analysis of low-density generator matrix codes with column weight 2 \cite{Hu}. 
Similar arguments were successfully used in Yano-Wadayama \cite{Yano} and Fujii-Wadayama \cite{Fujii} as well.

\section{Models and Definitions}


\subsection{Random Graph Model}

In this paper, we assume undirected $\lambda$-regular graphs with $n$ nodes, i.e., all the nodes have degree $\lambda$,
as a model of wireless packet-based networks.

Any undirected graph $\mathtt{G}$ can be converted into a bipartite graph $\mathtt{G}_{\mathrm{b}}$ 
by replacing each edge in $\mathtt{G}$ with a node of degree 2 and its connecting edges.
This bipartite graph $\mathtt{G}_{\mathrm{b}}$ is said to be the {\em counterpart} of $\mathtt{G}$. 
Conversely, $\mathtt{G}$ is also said to be the counterpart of $\mathtt{G}_{\mathrm{b}}$.
Then, the bipartite graph $\mathtt{G}_{\mathrm{b}}$ contains $n$ nodes of degree $\lambda$ and $(\lambda n)/2$ nodes of degree 2. 
By borrowing terms from the Tanner graph of LDPC codes, we refer the nodes of degree $\lambda$ (resp.\ 2) as variable (resp.\ check) nodes.
Summarizing above, we can convert any $\lambda$-regular graph into the counterpart, i.e., a $(\lambda, 2)$-regular bipartite graph.
Conversely, any $(\lambda, 2)$-regular bipartite graph can be converted into the $\lambda$-regular counterpart.

In this work, we will consider the robustness of networks over a random graph model 
based on a regular graph ensemble. 
We now define the regular graph ensemble associated with a regular bipartite graph ensemble, i.e., a regular Tanner graph ensemble.
For given parameters $n$ and $\lambda$, the $(\lambda,2)$-regular Tanner graph ensemble $\mathcal{T}(n,\lambda,2)$ is defined as follows \cite{modern}.
Any graph instance in $\mathcal{T}(n,\lambda,2)$ has $n$ variable nodes of degree $\lambda$ and $(\lambda n)/2$ check nodes of degree $2$.
A variable node of degree $\lambda$ has $\lambda$ sockets accepting edge connections from the check nodes.
Similarly, a check node of degree 2 has 2 sockets.
Hence, there are totally $\lambda n$ sockets in the both variable and check node sides.
Let $\pi$ be a permutation on $\{1,2,\dots,\lambda n\}$.
The $i$-th socket on the variable node side is connected to the $\pi(i)$-th socket on the check node side by an edge.
In the ensemble $\mathcal{T}(n,\lambda,2)$, we assume the uniform distribution on the set of permutations, i.e., 
the probability $1/(\lambda n) !$ is assigned to each bipartite graph.
The regular random graph ensemble $\mathcal{G}(n,\lambda)$ {\em is the counterpart of} $\mathcal{T}(n,\lambda,2)$.
More precisely, the graph ensemble $\mathcal{G}(n,\lambda)$ is the multiset of the counterparts of the bipartite graphs in $\mathcal{T}(n,\lambda,2)$
and the probability $1/(\lambda n)!$ is assigned to each element in $\mathcal{G}(n,\lambda)$, i.e., $\lambda$-regular graph.
Namely, there are one-to-one correspondence between $\mathcal{G}(n,\lambda)$  and $\mathcal{T}(n,\lambda,2)$.
This means that a $\lambda$ regular graph in $\mathcal{G}(n,\lambda)$ has the unique counterpart in $\mathcal{T}(n,\lambda,2)$ (vice versa) and the same probability is assigned to both graphs.
Some graphs in $\mathcal{G}(n,\lambda)$ contain self-loops or multiple edges but the fact causes no harmful effects on the following analysis.

\subsection{Node Fault Model \label{ssec:FM}}

Let $\mathtt{G} = (\mathtt{N}, \mathtt{E})$ be a $\lambda$-regular undirected graph.
We assume that each node in $\mathtt{N}$ independently breaks down with probability $\epsilon$.
Let $\mathtt{Z}$ be the set of broken nodes.
The {\em survivor graph} $\mathtt{G}\setminus \mathtt{Z}$ is the induced subgraph obtained 
by removing the broken nodes $\mathtt{Z}$ and their connected edges from the original graph $\mathtt{G}$.
If the survivor graph $\mathtt{G}\setminus \mathtt{Z}$ is {\em separated} (or {\em disconnected}), we consider the network {\em breakdown} happens because
there is at least a pair of nodes in $\mathtt{G}\setminus \mathtt{Z}$ such that they cannot communicate with each other.
In this paper, we consider the null graph is connected.
Hence, in the case that all the nodes in $\mathtt{G}$ are broken, 
we consider the network is {\em not broken}.

\section{Main Result}

In this section, we present the main result of this paper, i.e., 
an upper bound of average network breakdown probability under the node fault model defined in Subsection \ref{ssec:FM}. 

\subsection{Main Theorem} \label{main contribution}

Firstly, we give the notation to describe the theorem.
Let $\mathbb{Z}^{+}$ be the set of non-negative integers.
The indicator function $\mathbb{I}[cond]$ takes the value 1 if $cond$ is true, otherwise it gives the value 0.
We define the binomial coefficient as
\begin{equation*}
  {\textstyle
 \binom{n}{k} := 
  \frac{n!}{k!(n-k)!} \mathbb{I}[k, n-k \in \mathbb{Z}^+].}
\end{equation*}
Similarly, the multinomial coefficient is defined by
\begin{align*}
{\textstyle \binom{a_1+a_2 +\cdots +a_{k}}{a_1, a_2, \dots, a_k} 
  :=
  \frac{(a_1+a_2+\cdots +a_k)!}{a_1! a_2! \cdots a_k!} 
    \mathbb{I}[ a_1,a_2,\dots, a_k\in\mathbb{Z}^+].}
\end{align*}

The following theorem presents an upper bound of average network breakdown probability under the node fault model. 
\begin{theorem} \label{the:1}
Assume the node fault model defined in Subsection \ref{ssec:FM}.
The average network breakdown probability $P_{\lambda,n}(\epsilon)$,
whose expectation is taken over $\mathcal{G}(n,\lambda)$, 
is upper bounded by $P_{\lambda,n}^{(U)}(\epsilon)$, where
\begin{align}
&P_{\lambda,n}^{(U)}(\epsilon)
  :=
\sum_{j=0}^{n} {n \choose j}Q_{j,\lambda,n}^{(U)} \epsilon^j (1-\epsilon)^{n-j},
\\
 &Q_{j,\lambda,n}^{(U)}
  :=
 \frac{1}{2 \binom{\lambda n}{\lambda j}}
  \sum_{n_1=1}^{n-j-1}
   \frac{\binom{n-j}{n_1}}{  \binom{\lambda(n-j)}{\lambda n_1} } 
  \sum_{i_1=0}^{\lambda \min\{n_1, j\}} 
   2^{i_1} 
\notag \\
 &\qquad\quad\times
  \binom{\frac{\lambda n}{2}}
        {i_1, \frac{\lambda n_1 - i_1}{2}, \frac{\lambda (n-n_1) - i_1}{2}}
 \binom{\lambda (n-n_1) - i_1}{\lambda (n-n_1-j)}. 
 \label{eq:QU}
\end{align}
\end{theorem}

\begin{remark} \label{rem:1}
The value $Q_{j,\lambda,n}^{(U)}$ given in Theorem \ref{the:1} is an upper bound of $Q_{j,\lambda,n}$, where
$Q_{j,\lambda,n}$ is the probability that 
 a graph in $\mathcal{G}(n,\lambda)$ 
 become separated if a given set of $j$ nodes are broken.  
The precise definition of $Q_{j,\lambda,n}$ will be given  in \eqref{Qdef}.

We can easily verify that $Q_{n,\lambda,n}^{(U)} = 0$ from \eqref{eq:QU}.
This implies that the network is not broken down if all the nodes in $\mathtt{G}\in\mathcal{G}(n,\lambda)$ are broken.
Suppose that one wish to consider that a network breakdown happens if all the nodes in the graph are broken.
In such a case, by replacing $Q_{n,\lambda,n}^{(U)}$ with 1, one can easily obtain 
an upper bound of the average network breakdown probability.
\end{remark}

\subsection{Proof of Theorem \ref{the:1} \label{ssec:proof}}

\subsubsection{Ensemble average of network breakdown probability}

Let $\mathtt{G}=(\mathtt{N},\mathtt{E})$ be 
an undirected graph in $\mathcal{G}(n,\lambda)$.
 Recall that $\mathtt{G} \setminus \mathtt{Z}$ gives the survivor graph for $\mathtt{Z}\subset \mathtt{N}$.
Since each node is independently broken down with probability $\epsilon$,
the network breakdown probability $P_{\lambda,n}(\mathtt{G}, \epsilon)$ for $\mathtt{G}$
is given by
\begin{equation*} 
 P_{\lambda,n}(\mathtt{G}, \epsilon) 
  = 
 \sum_{j=0}^{n} \sum_{\mathtt{Z} \subset \mathtt{N}, |\mathtt{Z}| = j } 
 \mathbb{I}[\mathtt{G} \setminus \mathtt{Z}: \mbox{separated} ] 
\epsilon^j (1-\epsilon)^{n-j}.
\end{equation*}
By taking ensemble average over $\mathcal{G}(n,\lambda)$, we have
\begin{align}\nonumber
P_{\lambda,n}(\epsilon)  
&=  \sum_{j=0}^{n} \sum_{\mathtt{Z} \subset \mathtt{N}, |\mathtt{Z}| = j } 
{\sf E}[\mathbb{I}[\mathtt{G} \setminus \mathtt{Z}: \mbox{separated} ] ]
\epsilon^j (1-\epsilon)^{n-j} 
  \\ \nonumber
&= {\textstyle \sum_{j=0}^{n} \sum_{\mathtt{Z} \subset \mathtt{N}, |\mathtt{Z}| = j }} Q_{j,\lambda,n} \epsilon^j (1-\epsilon)^{n-j}
  \\ \label{symmetry}
&= {\textstyle \sum_{j=0}^{n} {n \choose j} } Q_{j,\lambda,n} \epsilon^j (1-\epsilon)^{n-j}.
\end{align}
The first equality is due to the linearity of expectation. 
The second equality directly follows from
the definition of $Q_{j,\lambda,n}$:
\begin{equation} \label{Qdef}
Q_{j,\lambda,n}
 := 
{\sf E}[ \mathbb{I} [\mathtt{G} \setminus \mathtt{Z}: \mbox{separated} ] ],
\end{equation}
where $\mathtt{Z}$ is any subset of $\mathtt{N}$ of size $j$.
Due to the symmetry of the ensemble, the right-hand side of \eqref{Qdef} depends only on the size of $\mathtt{Z}$ ($|\mathtt{Z}|=j$) instead of $\mathtt{Z}$ itself.
This fact leads to the last equality \eqref{symmetry}.

\subsubsection{Bipartite Configuration}

We further proceed the evaluation of $Q_{j,\lambda,n}$.
Let us fix any $\mathtt{Z} \subset \mathtt{V}$ ($|\mathtt{Z}| = j$).
Denote the counterpart of $\mathtt{G}\setminus \mathtt{Z}$, by $(\mathtt{G}\setminus \mathtt{Z})_{\mathrm{b}}$. 
The quantity $Q_{j,\lambda,n}$ can be rewritten as
\begin{align} \nonumber
Q_{j,\lambda,n} &= {\sf E}[ \mathbb{I} [\mathtt{G} \setminus \mathtt{Z}: \mbox{separated} ] ] 
 \\ \nonumber
&= {\textstyle \sum_{\mathtt{G} \in \mathcal{G}(n,\lambda) }} Prob(\mathtt{G}) \mathbb{I} [\mathtt{G} \setminus \mathtt{Z}: \mbox{separated} ] ] 
 \\ \label{enumeration}
&= 
{\textstyle \sum_{\mathtt{G}_{\mathrm{b}} \in \mathcal{T}(n,\lambda,2) } \mathbb{I} [(\mathtt{G} \setminus \mathtt{Z})_{\mathrm{b}}: \mbox{separated} ] ]} /(n \lambda)!. 
\end{align}
The third equality is due to the one-to-one correspondence between $\mathcal{G}(n,\lambda)$ and $\mathcal{T}(n,\lambda,2)$.  
The equation \eqref{enumeration} implies that 
the evaluation of $Q_{j,\lambda,n}$ can be reduced to the enumeration problem of
bipartite graphs in $\mathcal{T}(n,\lambda,2)$ that are separated if the nodes in $\mathtt{Z}$ are removed, 
namely, if the variable nodes corresponding to $\mathtt{Z}$ and the check nodes corresponding to the edges connecting to $\mathtt{Z}$ are removed.
The exact enumeration of such bipartite graphs is not a trivial problem. 
Instead of the exact enumeration,
we will derive an upper bound of the number of such graphs by counting bipartite configurations defined as follows.

\begin{figure}[tb]
   \centering
  \begin{picture}(150,82)
   \put(0,0){\includegraphics[width=150pt]{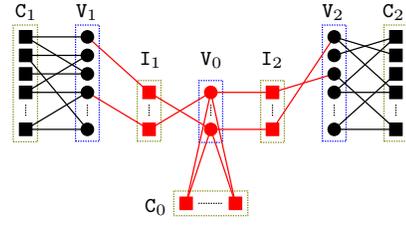}}
   \put(  1, 76){{\footnotesize $\mathtt{C}_1$}}
   \put( 24, 76){{\footnotesize $\mathtt{V}_1$}}
   \put( 48, 58){{\footnotesize $\mathtt{I}_1$}}
   \put( 71, 58){{\footnotesize $\mathtt{V}_0$}}
   \put( 94, 58){{\footnotesize $\mathtt{I}_2$}}
   \put(117, 76){{\footnotesize $\mathtt{V}_2$}}
   \put(140, 76){{\footnotesize $\mathtt{C}_2$}}
   \put( 50,  2){{\footnotesize $\mathtt{C}_0$}}
  \end{picture}
  \caption{ 
    Bipartite configuration.
    A variable (resp. check) node is depicted by a circle (resp. square).
    If all the red variable nodes $\mathtt{V}_0$, 
    the adjacent red check nodes $\mathtt{C}_0\cup\mathtt{I}_1\cup\mathtt{I}_2$,
    and their connecting red edges are removed from the graph, the remaining induced subgraph becomes separated.
}
  \label{fig:bg_f}
\end{figure}

Let $\mathtt{V}$ (resp.\ $\mathtt{C}$) be the set of variable (resp.\ check) nodes.
Assume that $\mathtt{V}$ and $\mathtt{C}$ 
are partitioned into $\mathtt{V} = \mathtt{V}_0 \cup \mathtt{V}_1 \cup \mathtt{V}_2$ and $\mathtt{C} = \mathtt{I}_1 \cup \mathtt{I}_2 \cup \mathtt{C}_0 \cup \mathtt{C}_1 \cup \mathtt{C}_2$, respectively. 
A tuple $\mathtt{F} = (\mathtt{V}_0, \mathtt{V}_1, \mathtt{V}_2, \mathtt{I}_1, \mathtt{I}_2, \mathtt{C}_0, \mathtt{C}_1, \mathtt{C}_2, 
\mathtt{E})$ is called a {\em bipartite configuration}
if each edge in $\mathtt{E}\subset \mathtt{V} \times \mathtt{C}$ satisfies the connection constraints shown in Fig.~\ref{fig:bg_f}.
In words, the variable nodes $\mathtt{V}_0$, $\mathtt{V}_1$, and $\mathtt{V}_2$ 
are adjacent to the check nodes 
$\mathtt{C}_0\cup\mathtt{I}_1\cup\mathtt{I}_2$, $\mathtt{C}_1\cup\mathtt{I}_1$, and $\mathtt{C}_2\cup\mathtt{I}_2$, respectively.
From this definition of the bipartite configuration,
it is clear that the graph $\mathtt{F}$ is bipartite.
This means that a bipartite configuration represents a bipartite graph
that are separated if all the variable nodes in $\mathtt{V}_0$ 
and the adjacent check nodes $\mathtt{C}_0\cup\mathtt{I}_1\cup\mathtt{I}_2$ are removed.
The bipartite graph $\mathtt{G}_{\mathrm{b}} = (\mathtt{V}\cup\mathtt{C},\mathtt{E})$ corresponding to the bipartite configuration $\mathtt{F}$ is said to be the {\em base graph} of $\mathtt{F}$.

\begin{figure}[tb]
  \begin{center}
  \begin{minipage}[t]{.48\linewidth}
    \centering
    \includegraphics[width=.9\linewidth]{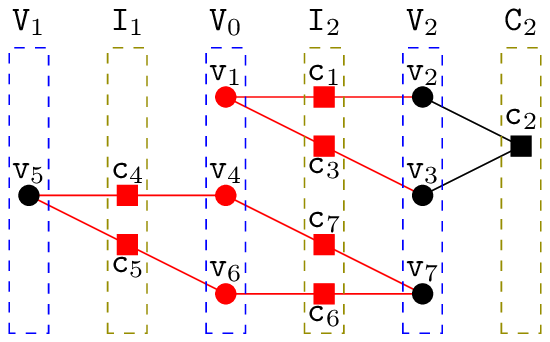}
    \subcaption{}
  \end{minipage}
  \begin{minipage}[t]{.48\linewidth}
    \centering
    \includegraphics[width=.9\linewidth]{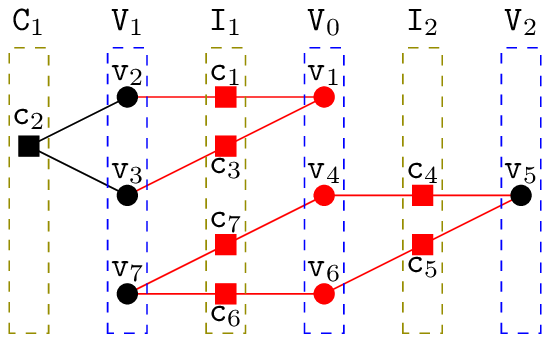}
    \subcaption{}    
  \end{minipage}
  \begin{minipage}[b]{.48\linewidth}
    \centering
    \includegraphics[width=.9\linewidth]{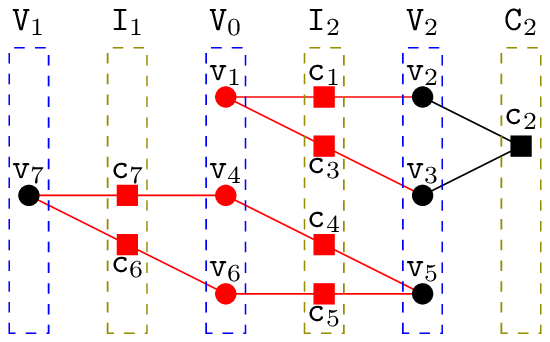}
    \subcaption{}    
  \end{minipage}
  \begin{minipage}[b]{.48\linewidth}
    \centering
    \includegraphics[width=.9\linewidth]{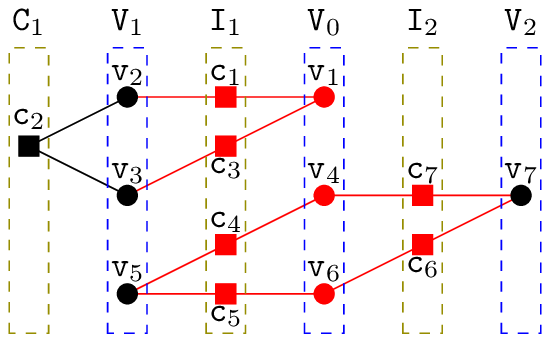}
    \subcaption{}    
  \end{minipage}
  \end{center}
  \caption{An example of four distinct bipartite configurations (a), (b), (c), (d) which have an identical base graph.}
  \label{fig:bc}  
\end{figure}

It should be remarked that two (or more) distinct bipartite configurations have an identical base graph.
Figure \ref{fig:bc} presents an example that four distinct bipartite configurations give the same base graph.
This {\em many-to-one correspondence} plays a crucial role to bound the number of bipartite graphs satisfying a certain condition.


\subsubsection{Enumeration of Bipartite Configurations}

We here enumerate the number of the bipartite configurations.
Define $n_k := |\mathtt{V}_k|$ and $c_k := |\mathtt{C}_k|$ for $k=0,1,2$, 
and let $i_k := |\mathtt{I}_k|$ for $k=1,2$.
Since the total number of the variable nodes is $n$, $n = n_0+n_1+n_2$ holds.
Since the sockets of $\mathtt{V}_1$ (resp.\ $\mathtt{V}_2$) are connected to the edges stemming 
from $\mathtt{C}_1$ (resp.\ $\mathtt{C}_2$) and $\mathtt{I}_1$ (resp.\ $\mathtt{I}_2$), 
we have $\lambda n_1 = 2 c_1 + i_1$ (resp.\ $\lambda n_2 = 2 c_2 + i_2$).
Similarly, we have $\lambda n_0 = 2 c_0 + i_1 + i_2$.
From the above constraints, if we fix $n_0, n_1, i_1, i_2$, 
then the size of $\mathtt{V}_2, \mathtt{C}_0, \mathtt{C}_1,\mathtt{C}_2$ are given by
\begin{equation}
 \begin{split}
 &n_2 = n-n_0-n_1,\quad c_0 = (\lambda n_0 - i_1 - i_2)/2, \\
 &c_1 = (\lambda n_1 - i_1)/2, \quad
  c_2 = \{\lambda (n-n_0-n_1) -i_2\}/2.
 \end{split}
 \label{eq:subs}
\end{equation}

For a given set of free parameters $n_0, n_1, i_1, i_2$, 
we denote the total number of the bipartite configurations with these parameters $n_0, n_1, i_1, i_2$, by $A_{n_0, n_1, i_1, i_2}$.
We will evaluate $A_{n_0, n_1, i_1, i_2}$ as follows.

The total numbers of assignments of variable nodes in $\mathtt{V}$ to $\mathtt{V}_0, \mathtt{V}_1, \mathtt{V}_2$
and check nodes $\mathtt{C}$ to $\mathtt{I}_1, \mathtt{I}_2, \mathtt{C}_0, \mathtt{C}_1, \mathtt{C}_2$
are given by $\binom{n}{n_0,n_1,n_2}$ and $\binom{\lambda n/2}{i_1, i_2, c_0, c_1, c_2}$, respectively.
For the sockets of $\mathtt{V}_1$ (resp.\ $\mathtt{V}_2$), 
$\binom{\lambda n_1}{i_1}$ (resp.\ $\binom{\lambda n_2}{i_2}$) gives the number of constellations of sockets which connect to the edges stemming from $\mathtt{I}_1$ (resp.\ $\mathtt{I}_2$).
Similarly, $\binom{\lambda n_0}{i_1, i_2, 2c_0}$ gives the number of socket constellations in $\mathtt{V}_0$.
Note that for every check node in $\mathtt{I}_1$ (resp.\ $\mathtt{I}_2$), one of its sockets connects to $\mathtt{V}_0$ and the other socket connects to $\mathtt{V}_1$ (reps.\ $\mathtt{V}_2$).
Hence, the number of socket constellations in $\mathtt{I}_1$ (resp.\ $\mathtt{I}_2$) is $2^{i_1}$ (resp.\ $2^{i_2}$).
The total number of the edge permutations is given by $(2 c_1)! (2 c_2)! (i_1!)^2 (i_2!)^2 (2i_0)!$.
By multiplying these numbers, we immediately obtain
\begin{align}
  A_{n_0, n_1, i_1, i_2} 
 =
 (\lambda n)! \frac{\binom{n}{n_0, n_1, n_2}\binom{\lambda n/2}{i_1, i_2, c_0, c_1, c_2}}{\binom{\lambda n}{\lambda n_0, \lambda n_1, \lambda n_2}} 
 2^{i_1+i_2}.
 \label{eq:A}
\end{align}

\subsubsection{Derivation of an Upper Bound of $Q_{j,\lambda,n}$}

Recall that the number of the graphs that become separated by removing $\mathtt{Z}$ leads to an upper bound of $Q_{j,\lambda,n}$. 
We here prepare a tool to evaluate the number of such graphs.
Let $K(\mathtt{W})$ $(\mathtt{W} \subset \mathtt{V})$ be the number of the bipartite configurations satisfying 
$\mathtt{V}_0 = \mathtt{W}$ $(|\mathtt{W}| = j)$. 
From the definition of $A_{j, n_1, i_1, i_2}$ and the reasons described in later, $K({\mathtt W})$ is given by
\begin{equation}\label{KZ}
 K({\mathtt W}) 
  = 
 \frac{1}{{n \choose j} }
 \sum_{n_1 = 1}^{n-j-1} \sum_{i_1 = 0}^{\lambda \min\{j,n_1\} } 
 \sum_{i_2 = 0}^{\min \{\lambda n_2, \lambda j - i_1\}}
  A_{j, n_1, i_1, i_2}.
\end{equation}
The factor $1/{n \choose j}$ is introduced because the triple summations in the right-hand side count the number of 
the bipartite configurations to be separated if {\em any} set of $j$ nodes are removed.
Since the right-hand side depend only on $j$, we will denote $K(\mathtt{W})$  by $K(j)$.

The upper and lower bounds of $n_1, i_1, i_2$ in \eqref{KZ} are determined as follows.
Even for $i_1 = 0$ or $i_2 = 0$, the survivor graph becomes separated if the nodes in $\mathtt{V}_0\cup\mathtt{C}_0\cup\mathtt{I}_1\cup\mathtt{I}_2$ are removed.
Hence, $i_1\ge 0$ and $i_2 \ge 0$ holds.
The size of $\mathtt{I}_1$ is upper bounded by the number of connecting edges of $\mathtt{V}_0$ and $\mathtt{V}_1$.
This fact leads to the inequality $0 \le i_1 \le \lambda \min\{j, n_1\}$.
Similarly, for a fixed $i_1$, the size of $\mathtt{I}_2$ is upper bounded 
by the number of connecting edges of $\mathtt{V}_2$ and the number of remaining edges connecting to $\mathtt{V}_0$.
This gives the inequalities  $0 \le i_2 \le \min\{\lambda j - i_1, \lambda n_2\}$.
A survivor graph becomes separated if $n_1 \ge 1$ and $n_2 \ge 1$ when the nodes in $\mathtt{V}_0\cup\mathtt{C}_0\cup\mathtt{I}_1\cup\mathtt{I}_2$ are removed.
We thus have $1 \le n_1 \le n-j-1$.

For a given $\mathtt{W}\subset \mathtt{V}$ such that $|\mathtt{W}| = j$,
let $\mathcal{F}$ be the set of bipartite configurations defined by
$
\mathcal{F} := \{\mathtt{F} \mid \mathtt{V}_0 = \mathtt{W} \}.
$
Assume that we have the base graph $\mathtt{G}_{\mathrm{b}}$ corresponding to $\mathtt{F} \in \mathcal{F}$.
By removing $\mathtt{V}_0 \cup \mathtt{C}_0\cup\mathtt{I}_1\cup\mathtt{I}_2$ and the connecting edges from the bipartite graph $\mathtt{G}_{\mathrm{b}}$,  
the remaining induced subgraph becomes separated (see also Fig.~\ref{fig:bg_f}).
Due to the many-to-one correspondence between bipartite configurations 
and base graphs, we immediately have the following upper bound:
\begin{align*} 
{\textstyle \sum_{\mathtt{G}_{\mathrm{b}} \in \mathcal{T}(n,\lambda,2) }  }
\mathbb{I}[(\mathtt{G} \setminus \mathtt{Z})_{\mathrm{b}}: \mbox{separated} ] ] 
\le | \mathcal{F} |/2  = K(j)/2,
\end{align*}
for a given $\mathtt{Z} \subset \mathtt{N}$ such that $|\mathtt{Z}| = j$.
The factor $1/2$ compensates the over counting due to the trivial many-to-one correspondence,
i.e., the swap of the left-hand side $(\mathtt{C}_1, \mathtt{V}_1, \mathtt{I}_1)$ and the right-hand side $(\mathtt{C}_2, \mathtt{V}_2, \mathtt{I}_2)$ 
in Fig.~\ref{fig:bg_f}.
From the above argument and \eqref{enumeration}, we get the bound of $Q_{j,\lambda,n}$:
\begin{equation} \label{Qbound}
 Q_{j,\lambda,n} 
  \le
 K(j) / \{2 (\lambda n)!\}.
\end{equation}
From \eqref{eq:subs}, \eqref{eq:A} and the definition $K(j)$, we have
\begin{align*}
  &K(j)/\{2 (\lambda n)!\}
\\
  =&
 \frac{1}{ 2 \binom{\lambda n}{\lambda j} }
 \sum_{n_1=1}^{n-j-1}\frac{\binom{n-j}{n_1}}{ \binom{\lambda(n-j)}{\lambda n_1} } 
\sum_{i_1 = 0}^{\lambda \min\{n_1,j\} }\! 
  \binom{\frac{\lambda n}{2}}
   {i_1, \frac{\lambda n_1 - i_1}{2}, \frac{\lambda (n- n_1) - i_1}{2}} 
  \\
 &\!\!\!\!\times  
  2^{i_1}  \sum_{i_2 = 0}^{\min \{\lambda (n-n_1-j), \lambda j - i_1\}} \!
  \binom{\frac{\lambda (n- n_1) - i_1}{2}}
   {i_2, \frac{\lambda j - i_1 -i_2}{2}, \frac{\lambda (n-j-n_1) -i_2}{2}}
  2^{i_2}.
\end{align*}
The identity on the following sum\footnote{
This identity can be derived in a similar way to \cite[Exercises 5.24]{Graham_Concrete}.}
\begin{align*}
 \sum_{i=0}^{\min\{a,b\}} 
 \binom{\frac{a+b}{2}}{\frac{a-i}{2}, \frac{b-i}{2}, i}2^{i} 
 =
 \begin{cases}
  \binom{a+b}{a}, & \text{if}~~\frac{a+b}{2}\in \mathbb{Z}^{+}, \\
  0, & \text{otherwise},
 \end{cases}
\end{align*}
leads to the identity:
\begin{equation} \label{QU}
 K(j)/\{2 (n \lambda)!\}
 =
 Q_{j,\lambda,n}^{(U)}.
\end{equation}
By combining \eqref{symmetry}, \eqref{Qbound} and \eqref{QU},  
we finally obtain the claim of the theorem.

\subsection{Computer Experiments \label{ssec:CE}}

The details of the experiments are as follows.
One experiment includes $o_{max}$-unit experimental trials.
A unit experimental trial consists of two phases: generation of a network instance and depth first search (DFS) processes for examining the connectivity of survivor graphs.
In the first phase of a trial, 
we randomly pick up a bipartite graph $\mathtt{G}_{\mathrm{b}}$ from the ensemble $\mathcal{T}(n,\lambda,2)$ and then
we construct its counterpart $\mathtt{G}$, i.e., an undirected $\lambda$-regular graph representing a network.
Let $i_{max}$ be the number of iterations for the DFS processes. 
For a fixed $\mathtt{G}$, the second phase consists of $i_{max}$-times executions of the DFS processes.
For each DFS process, some of the nodes are destroyed according to the node fault model with the given probability $\epsilon$.
The connectivity of the survivor graph is checked by using a simple DFS.
This means that $i_{max}$-survivor networks are checked for a fixed graph $\mathtt{G}$. 
During the experimental trials, we can count the number of network breakdown events and it gives an estimate of the average network breakdown probability.
In the following experiments, we used the parameters $o_{max} = 10^4$ and $i_{max} = 10^5$.

\begin{figure}[tbp]
  \centering
    \includegraphics[scale=0.45]{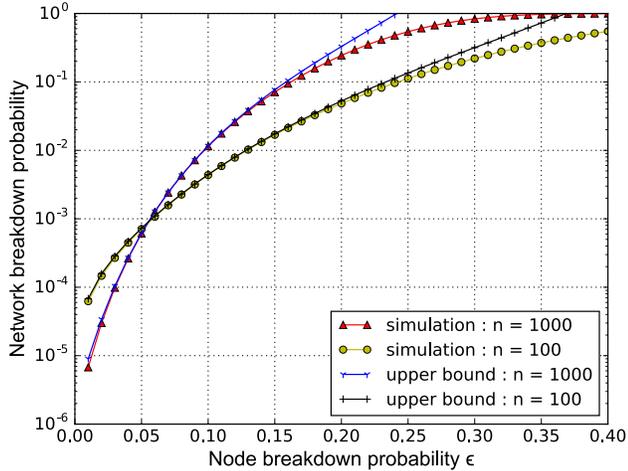}
  \caption
{
Comparisons of network breakdown probabilities obtained by computer experiments and upper bounds (Theorem \ref{the:1}). 
$\lambda = 5$, $n = 100, 1000$.
}
\label{fig:failure_prob1}
\end{figure}

\begin{figure}[tbp]
  \centering
    \includegraphics[scale=0.45]{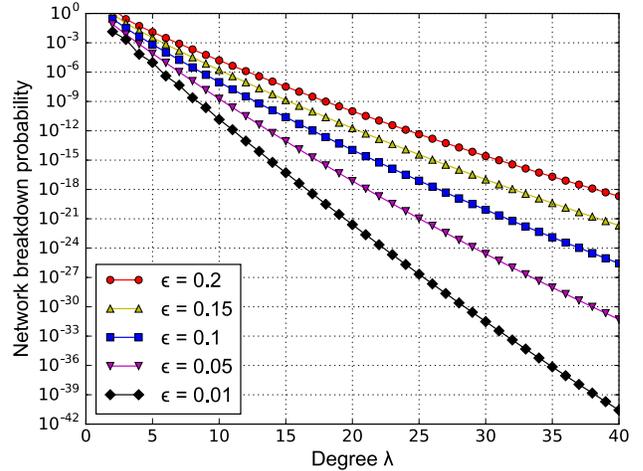}
  \caption
{
Upper bounds of network breakdown probabilities as a function of the degree $\lambda$ ($n=100$).
}
\label{fig:failure_prob2}
\end{figure}

Figure \ref{fig:failure_prob1} presents 
the average network breakdown probabilities estimated by the computer experiments in the case of $n = 100, 1000$ and $\lambda = 5$.
The horizontal axis represents the node breakdown probability $\epsilon$.  
The figure also includes the values of upper bound in Theorem \ref{the:1} as a function of $\epsilon$.
We can observe that the values of the estimated average and the upper bound are very close in both cases (i.e., $n = 100, 1000$) especially when $\epsilon$ is less than 0.15.
This result provides an evidence that the upper bound in Theorem \ref{the:1} is tight if $\epsilon$ is small enough.
Figure \ref{fig:failure_prob2} shows a relationship between the upper bound of average network breakdown probability and the degree $\lambda$. 
As intuition tells us, the average network breakdown probability is increased when the degree $\lambda$ gets larger.
From Fig.~\ref{fig:failure_prob2}, we can observe that the average network breakdown probability appears an exponential function of the degree $\lambda$.

\section{Concluding Summary}

In the present paper, we have derived an upper bound of the average network breakdown probability of packet networks with unreliable relay nodes.
Experimental results indicate that the upper bound is extremely tight when $\epsilon$ is small.
This implies that the over counting introduced in the proof of the bound becomes negligible if $\epsilon$ is small.
The random geometric graph models are often used to model WSNs. 
It is known that the degree distribution of a random geometric graph becomes tightly concentrated to the average degree of it when the number of nodes are large enough. 
We might be able to expect that our bound gives reasonable approximations in such cases as well.

From the upper bound and the experimental results, we 
have observed that the average network breakdown probability behaves as an exponential function of the degree $\lambda$. 
The degree $\lambda$ is a local quantity that indicates local connectivity of the graph.
On the other hand, the average network breakdown probability is a global quantity that indicates the robustness of the networks 
against possible breakdowns of unreliable relay nodes.
It is interesting to pursue a rigorous argument on this relationship in the asymptotic regime.


\section*{Acknowledgment}
This work is supported by JSPS Grant-in-Aid for Scientific Research
16K16007 and 16H02878.

\bibliographystyle{IEEEtran}
\bibliography{IEEEabrv,bib_isit2017}

\begin{thebibliography}{10}
\providecommand{\url}[1]{#1}
\csname url@samestyle\endcsname
\providecommand{\newblock}{\relax}
\providecommand{\bibinfo}[2]{#2}
\providecommand{\BIBentrySTDinterwordspacing}{\spaceskip=0pt\relax}
\providecommand{\BIBentryALTinterwordstretchfactor}{4}
\providecommand{\BIBentryALTinterwordspacing}{\spaceskip=\fontdimen2\font plus
\BIBentryALTinterwordstretchfactor\fontdimen3\font minus
  \fontdimen4\font\relax}
\providecommand{\BIBforeignlanguage}[2]{{%
\expandafter\ifx\csname l@#1\endcsname\relax
\typeout{** WARNING: IEEEtran.bst: No hyphenation pattern has been}%
\typeout{** loaded for the language `#1'. Using the pattern for}%
\typeout{** the default language instead.}%
\else
\language=\csname l@#1\endcsname
\fi
#2}}
\providecommand{\BIBdecl}{\relax}
\BIBdecl

\bibitem{Li}
J.~Li, L.~L. Andrew, C.~H. Foh, M.~Zukerman, and H.-H. Chen, ``Connectivity,
  coverage and placement in wireless sensor networks,'' \emph{Sensors}, vol.~9,
  no.~10, pp. 7664--7693, 2009.

\bibitem{REL}
E.~F. Moore and C.~E. Shannon, ``Reliable circuits using less reliable
  relays,'' \emph{Journal of the Franklin Institute}, vol. 262, no.~3, pp.
  191--208, 1956.

\bibitem{NP1}
J.~S. Provan and M.~O. Ball, ``The complexity of counting cuts and of computing
  the probability that a graph is connected,'' \emph{SIAM Journal on
  Computing}, vol.~12, no.~4, pp. 777--788, 1983.

\bibitem{NP2}
L.~G. Valiant, ``The complexity of enumeration and reliability problems,''
  \emph{SIAM Journal on Computing}, vol.~8, no.~3, pp. 410--421, 1979.

\bibitem{Poly}
D.~R. Karger and R.~P. Tai, ``Implementing a fully polynomial time
  approximation scheme for all terminal network reliability,'' in
  \emph{Proceedings of the eighth annual ACM-SIAM symposium on Discrete
  algorithms}.\hskip 1em plus 0.5em minus 0.4em\relax Society for Industrial
  and Applied Mathematics, 1997, pp. 334--343.

\bibitem{modern}
T.~Richardson and R.~Urbanke, \emph{Modern coding theory}.\hskip 1em plus 0.5em
  minus 0.4em\relax Cambridge University Press, 2008.

\bibitem{Hu}
C.-H. Hsu and A.~Anastasopoulos, ``Capacity-achieving codes with bounded
  graphical complexity and maximum likelihood decoding,'' \emph{IEEE
  Transactions on Information Theory}, vol.~56, no.~3, pp. 992--1006, 2010.

\bibitem{Yano}
A.~Yano and T.~Wadayama, ``Probabilistic analysis of the network reliability
  problem on random graph ensembles,'' \emph{IEICE Transactions on Fundamentals
  of Electronics, Communications and Computer Sciences}, vol. E99-A, no.~12,
  pp. 2218--2225, 2016.

\bibitem{Fujii}
Y.~Fujii and T.~Wadayama, ``Probabilistic analysis on minimum st cut capacity
  of random graphs with specified degree distribution,'' \emph{IEICE
  Transactions on Fundamentals of Electronics, Communications and Computer
  Sciences}, vol. 97-A, no.~12, pp. 2317--2324, 2014.

\bibitem{Graham_Concrete}
R.~L. Graham, D.~E. Knuth, and O.~Patashnik, \emph{Concrete Mathematics: A
  Foundation for Computer Science}, 2nd~ed.\hskip 1em plus 0.5em minus
  0.4em\relax Addison-Wesley Longman Publishing Co., Inc., 1994.

\end{thebibliography}

\end{document}